\documentclass[10pt]{article}
\usepackage{graphicx}

\def\Title#1{\begin{center} {\Large #1} \end{center}}
\def\Author#1{\begin{center}{ \sc #1} \end{center}}
\def\Address#1{\begin{center}{ \it #1} \end{center}}

\newcommand\pubblock{\rightline{\begin{tabular}{l} Proceedings of the Second Annual LHCP\\ \pubnumber\\
         \pubdate  \end{tabular}}}

\newenvironment{Abstract}{\begin{quotation} \begin{center} 
             \large ABSTRACT \end{center}\bigskip 
      \begin{center}\begin{large}}{\end{large}\end{center} \end{quotation}}

\newenvironment{Presented}{\begin{quotation} \begin{center} 
             PRESENTED AT\end{center}\bigskip 
      \begin{center}\begin{large}}{\end{large}\end{center} \end{quotation}}

\def\beq{\begin{equation}}
\def\eeq#1{\label{#1}\end{equation}}
\def\eeqn{\end{equation}}

\def\beqa{\begin{eqnarray}}
\def\eeqa#1{\label{#1}\end{eqnarray}}
\def\eeqan{\end{eqnarray}}

\let\bar=\overbar

\def\Dslash{\not{\hbox{\kern-4pt $D$}}}
\def\dslash{\not{\hbox{\kern-2pt $\del$}}}

\def\msb{{\bar{\ssstyle M \kern -1pt S}}}

\usepackage{color}

\newcommand\pubnumber{ CDF/PUB/BOTTOM/PUBLIC/11118 }

\newcommand\pubdate{\today}

\def\affiliation{
On behalf of the CDF and D0 Experiments, \\
Istituto Nazionale di Fisica Nucleare - Sezione di Pisa \\
Largo B. Pontecorvo, 3, Pisa, I 56127, Italy }

\begin{document}

\large
\begin{titlepage}
\pubblock

\vfill
\Title{  Tevatron Results on Heavy Flavor Production and Decays }
\vfill

\Author{ Fabrizio SCURI  }
\Address{\affiliation}
\vfill
\begin{Abstract}

The most recent results on heavy flavor production and decays from the Tevatron experiments CDF and 
D0 are summarized and compared with some LHC experiment results. The collected data sample refers to
the full Tevatron run-II operation and it corresponds to about 10 fb$^{-1}$ of integrated luminosity
per experiment. 

\end{Abstract}
\vfill

\begin{Presented}
The Second Annual Conference\\
 on Large Hadron Collider Physics \\
Columbia University, New York, U.S.A \\ 
June 2-7, 2014
\end{Presented}
\vfill
\end{titlepage}
\def\thefootnote{\fnsymbol{footnote}}
\setcounter{footnote}{0}
%

\begin{normalsize} 


\section{Introduction}

A wide heavy flavor physics program was done during the Tevatron Run II data taking thanks to the
large b-hadron cross-section production, $\sigma(p\bar{p}\rightarrow b\bar{b}) \simeq 50 ~\mu b$ at 
$\sqrt{s} = 2$ TeV, which is a factor $10^{4}-10^{5}$ larger than the production cross-section 
$\sigma(e^{+}e^{-}\rightarrow  b\bar{b})$ at the B-factories [Y(4s)] active in the same decade.

All b-hadrons ($B^{+},B^{0},B_{s},B_{c},\Lambda_{b},\Sigma_{b},\Xi_{b},\Omega_{b}$) are produced in 
$p\bar{p}$ collisions at Tevatron with production fractions $f_{d}:f_{u}:f_{s}:f_{\Lambda} \sim 4:4:1:1$, 
allowing a physics program complementary to the one of the B-Factories; however, the inelastic 
cross-section $\sigma(p\bar{p})_{inel.} \sim 100 ~mb$ is a factor $10^{3}-10^{4}$ larger than
$\sigma(p\bar{p}\rightarrow b\bar{b})$, and the branching ratios of rare b-hadron decays are
$O(10^{-6})$ or lower; therefore, in order to efficiently reconstruct b-events, detectors need to have 
a very good tracking and vertex resolution, a wide acceptance and good particle identification for 
electrons and muons, and a highly selective trigger. The complete description of the CDF and D0 
detectors can be found elsewhere \cite{xxx:detectors}. Different trigger 
types are used at the Tevatron experiments to select events with b-hadron production; both CDF and D0 
use dimuon ($J/\Psi$ modes) triggers with $p_{t}(\mu) > 1.5 ~GeV$, while CDF exploits also a special 
tool for secondary vertex reconstruction, the silicon vertex tracker (SVT) trigger, which reconstructs 
tracks displaced with respect to the primary vertex and having $p_{t} > 2 ~GeV/c$ and impact parameter
$d_{0} > 100 ~\mu m$ \cite{xxz:cdf-svt}; with this trigger type fully hadronic b-decay modes can be 
also reconstructed.

Some recent results from the Tevatron experiments are presented here for b-hadrons;
the analyzed data are relative to the full Tevatron Run II sample, corresponding to about 10 fb$^{-1}$
of integrated luminosity per experiment. Results are grouped as it follows: \\
- section 2: B-mesons. 

- section 2.1: rare decays: $B_{s} \rightarrow \mu^{+} \mu^{-}$. 

- section 2.2: $B_{c}$ semileptonic decays: $B^{+}_{c}\rightarrow J/\Psi \mu^{+}\nu$.
 
- section 2.3: orbitally excited B-mesons ($B^{0,+}_{1}, B^{*0,+}_{2}, B^{0}_{s1}, B^{*0}_{s2}$), and new $B\pi$ 
resonances: $B(5970)^{0,+}$. \\ 
- section 3: b baryons: $\Xi^{0,-}_{b}, \Omega^{-}_{b}$. \\
- section 4: Exotic resonances: $X(4140)$.   

\section{B-mesons}
\subsection{Rare decays: $B_{s} \rightarrow \mu^{+} \mu^{-}$}

The long study of the $B_{s} \rightarrow \mu^{+} \mu^{-}$ decay is a clear example of a rich legacy 
left by Tevatron to the present and future hadron colliders; in more than 10 years, a variety of
methods and tools was developed, having outlined the main road for the recent LHC evidence results
\cite{xyx:lhc-bsmumu}.
\begin{figure}[htb]
\begin{center}
\begin{tabular}{lll}
\includegraphics[height=1.7in]{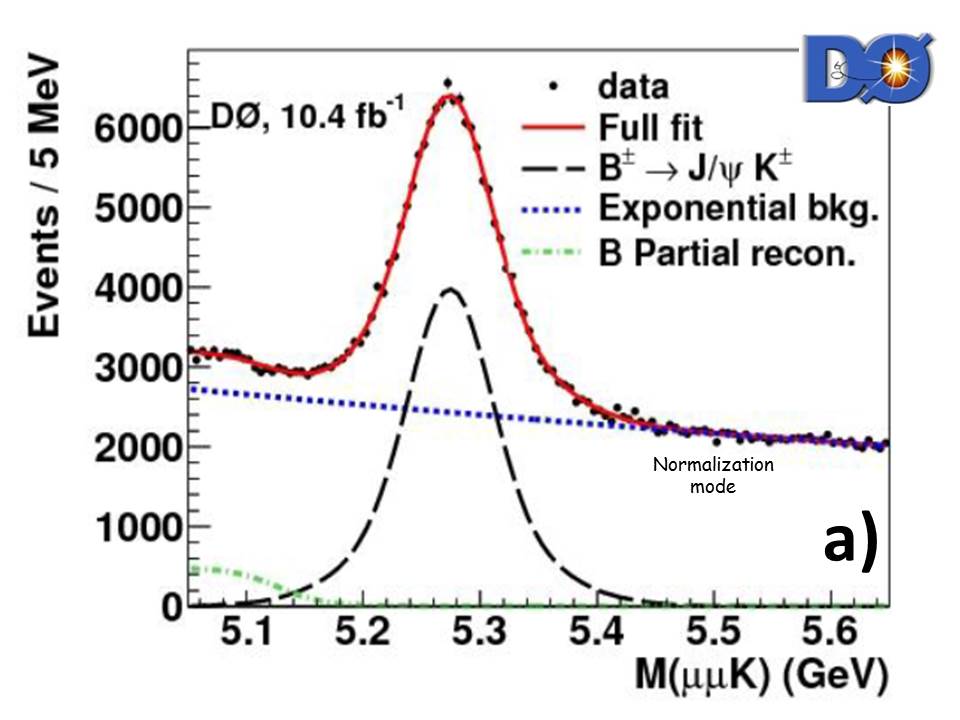} &
\includegraphics[height=1.7in]{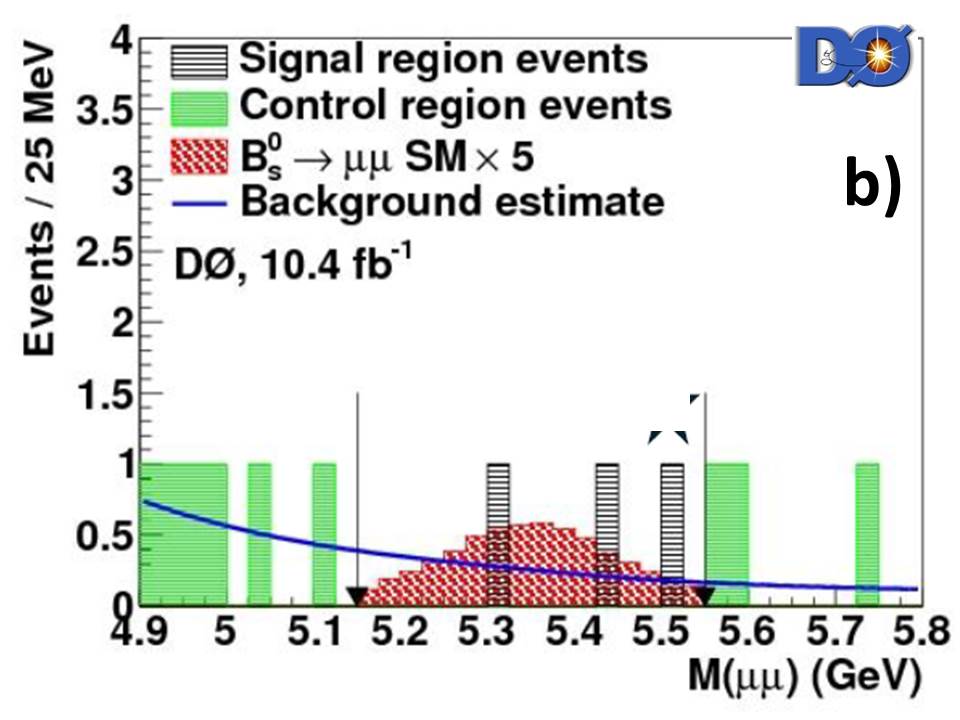} &
\includegraphics[height=1.7in]{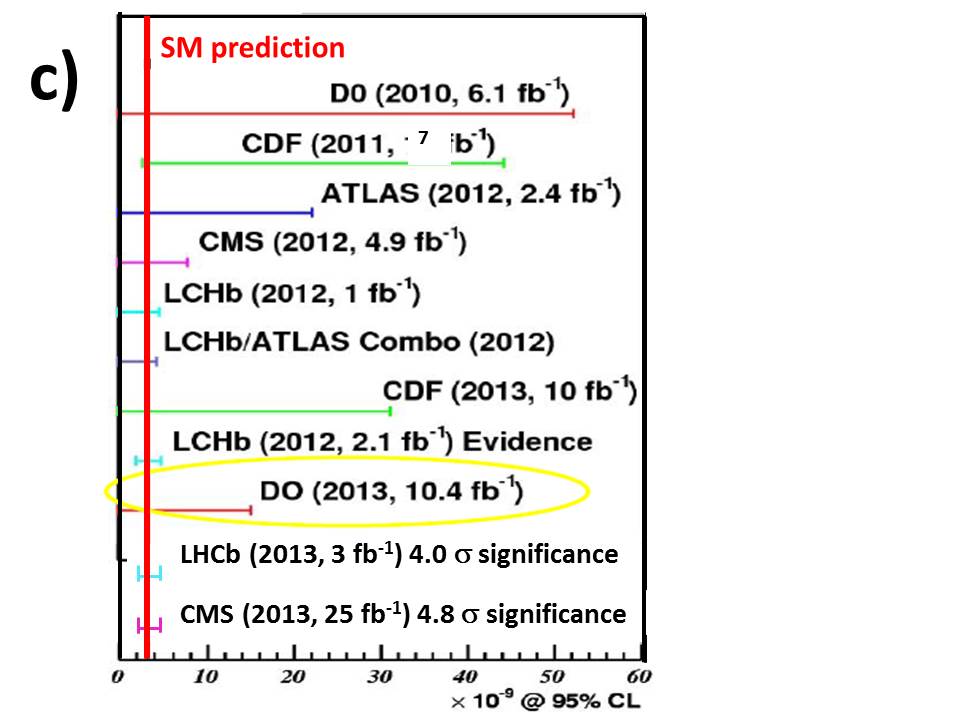}
\end{tabular}
\caption{ {\footnotesize a): Normalization mode invariant mass; b): M($\mu \mu$) invariant mass after all selection 
cuts; c): Summary of the latest $BR(B^{0}_{s} \rightarrow \mu \mu)$ measurements.} }
\label{fig:fig1}
\end{center}
\end{figure}  

High purity and efficiency selection of dimuon vertex candidates is the first step of the procedure;
efficient rejection of the background is obtained by applying multivariate analysis techniques such
as Neural Network (NN) and Boosted Decision Tree (BDT); the kinematic observables of the tracks and
of the reconstructed vertexes as well as global event information are considered to select the event
candidate sample and to deplete the background contributions (mainly $B \rightarrow h^{+}h^{-}$ decays
and partially reconstructed B decays); the background level in the signal region is extrapolated from
the sideband control regions; Single Event Sensitivity (SES) for the signal is determined from the 
abundant normalization mode $B^{+} \rightarrow J/\Psi K^{+} \rightarrow (\mu^{+}\mu^{-}) K^{+}$; Bayesian 
and frequentist approaches are used to set the expected (from SES) and observed limits at 
90$\%$(95$\%$) confidence level. Details of the individual analyses can be found in the last Tevatron
publications \cite{xyz:tev-bsmumu}.  

D0 set the best Tevatron limits: $BR(B^{0}_{s} \rightarrow \mu^{+} \mu^{-}) < 15 \times 10^{-9} (12 
\times 10^{-9})$ at $90\% ~(95 \%)$ confidence level, only a factor 5 above the value predicted by 
the Standard Model and recently confirmed  by the LHC experiment evidences. The invariant mass spectra of the 
normalization channel and of dimuon mass in the signal region are shown in figure \ref{fig:fig1}a) and 
b) respectively, while the summary of most recent world results are shown in figure \ref{fig:fig1}c).

\subsection{$B_{c}$ semileptonic decays: $B^{+}_{c}\rightarrow J/\Psi \mu^{+}\nu$}

The $B^{+}_{c}$ meson is the most massive bottom-flavored meson, apart from $b\bar{b}$, and it consists of a 
$\bar{b}$ quark and a $c$ quark in the ground state; it was discovered by CDF in the Tevatron Run I and it is
an unique laboratory to study QCD and weak decays. The dominant production in $p\bar{p}$ collisions is through
hard processes (figure \ref{fig:fig5}a)) and the $B^{+}$ meson decays only weakly; the decay modes with 
$b$-quark spectator and $c$-quark spectator have different final states, so they do not interfere.

The CDF experiment has recently completed an analysis on the full Run II data set corresponding to 8.7 fb$^{-1}$
of integrated luminosity; the measured observable is the $B^{+}_{c}$ production cross/section times the 
branching ratio in the $J/\Psi \mu^{+} \nu $ decay mode, normalized to the same quantity for the normalization mode
$B^{+} \rightarrow J/\Psi K^{+}$, and using the dimuon trigger type. Event selection is based on the 
association to the $J/\Psi$ vertex of a third track that may be: the muon in the $B^{+}_{c} \rightarrow 
J/\Psi \mu^{+} X$ decays; a charged kaon in the $B^{+} \rightarrow J/\Psi K^{+}$ decay; a $\pi^{+} , K^{+}$, or
$p$ in the background control samples. The invariant mass distribution of the $B^{+}_{c}\rightarrow J/\Psi 
\mu^{+}\nu$ candidate events is shown if figure \ref{fig:fig5}b).
\begin{figure}[htb]
\begin{center}
\begin{tabular}{cc}
\includegraphics[height=1.7in]{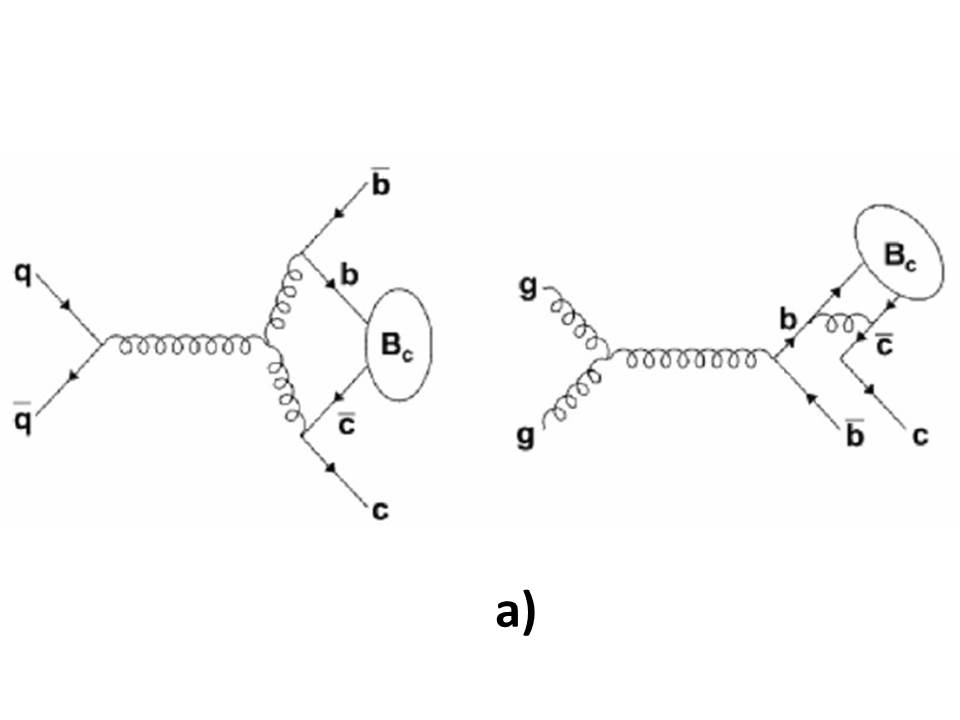} &
\includegraphics[height=1.7in]{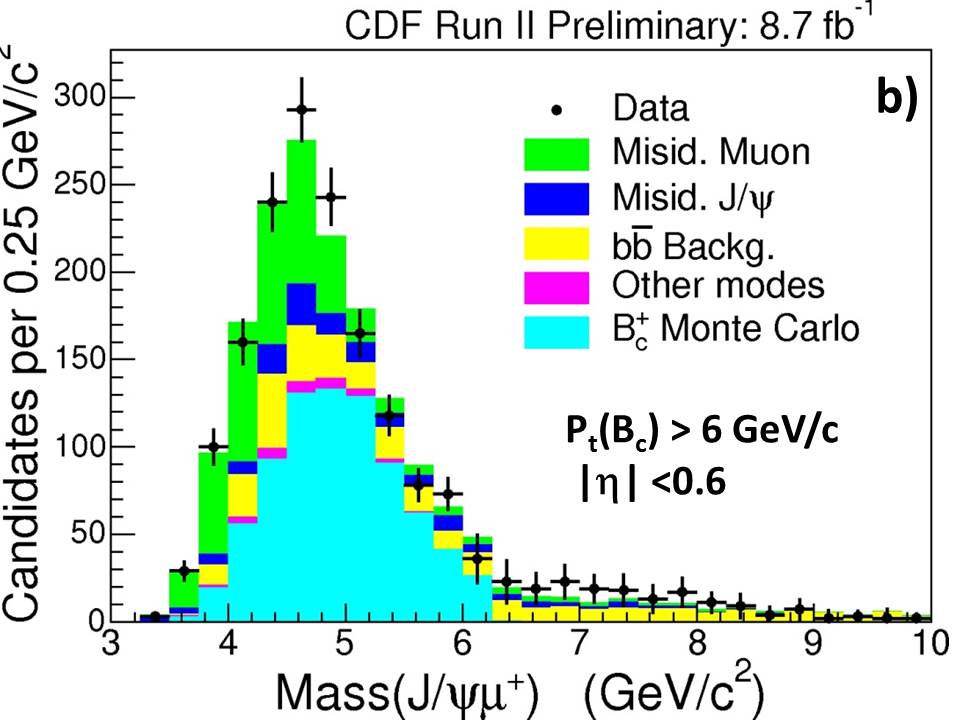} 
\end{tabular}
\caption{{\footnotesize  a): examples of $B_{c}$ production hard processes; b): invariant mass of the 
$B^{+}_{c}\rightarrow J/\Psi \mu^{+}\nu$ candidate events.}}
\label{fig:fig5}
\end{center}
\end{figure} 

The new CDF results is:
\begin{equation}
\frac{\sigma(B^{+}_{c}) \times BR(B^{+}_{c} \rightarrow J/\Psi \mu^{+} \nu)}{\sigma(B^{+}) \times BR(B^{+}
 \rightarrow J/\Psi K^{+})} = 0.211 \pm 0.012 (stat.)^{+0.021}_{-0.020}(syst.);
\end{equation}
the systematic error is dominated by the uncertainty on muon identification and the muon efficiency \cite{xxy:bc}. 

\subsection{Orbitally excited B-mesons and new resonances}

The study of the properties of orbitally excited B-mesons allows accurate tests of the predictions of 
the heavy-quark effective theory (HQET); assuming the bottom quark to be heavy, like the proton in the 
hydrogen atom, the dynamics in HQET is dominated by the coupling between the light-quark's orbital momentum
and spin, resulting in the total light-quark's momentum $j$; additional contributions to 
system dynamics arise due to the coupling between the b-quark spin and $j$; this results in two doublet 
states, corresponding to the fine and hyperfine splitting shown in figure \ref{fig:fig2}a). Two states
($B_{1}$ and $B^{*}_{2}$) are $narrow$ due to parity and angular momentum conservation; three decays
per flavor mode ($B_{1(s1)} \rightarrow B^{*}\pi (K), ~B^{*}_{2(s2)} \rightarrow B^{*}\pi (K)$, and 
$B^{*}_{2(s2)} \rightarrow B\pi (K)$) are observed by CDF; the other states have predicted widths of 150 
$MeV/c^{2}$ and are too broad to be detected. The sum of the individual samples with a kaon and 
including B decays in the $J/\Psi$ mode and in fully hadronic modes with a D-meson is shown in 
figure \ref{fig:fig2}b); signals are described by non-relativistic Breit-Wigner functions convoluted 
with two Gaussians to account for the detector resolution.
   
\begin{figure}[htb]
\begin{center}
\begin{tabular}{cc}
\includegraphics[height=1.7in]{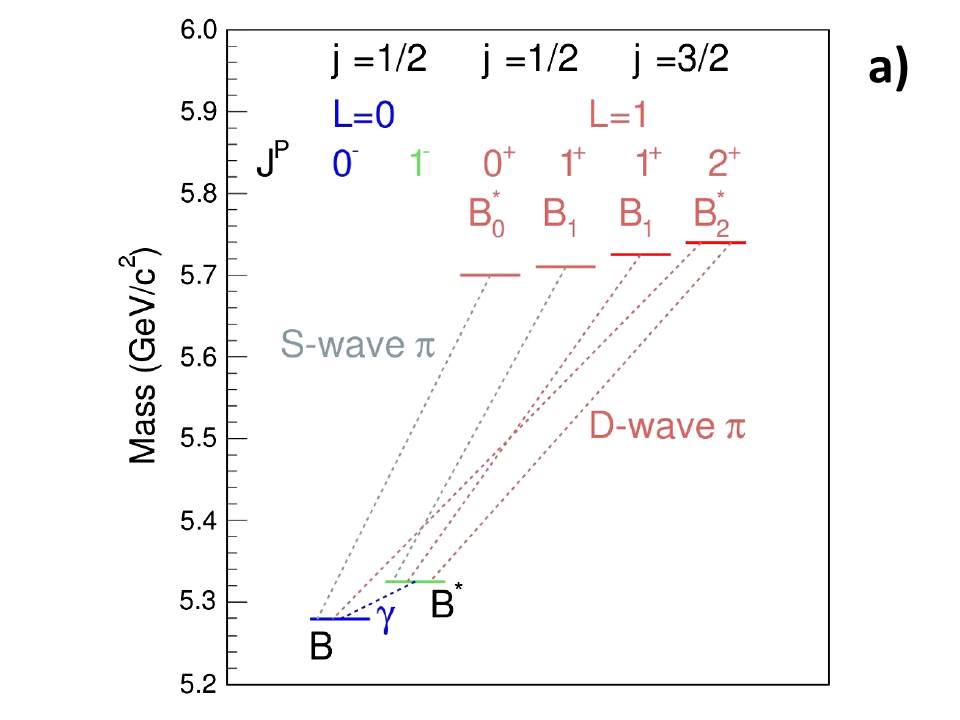} &
\includegraphics[height=1.7in]{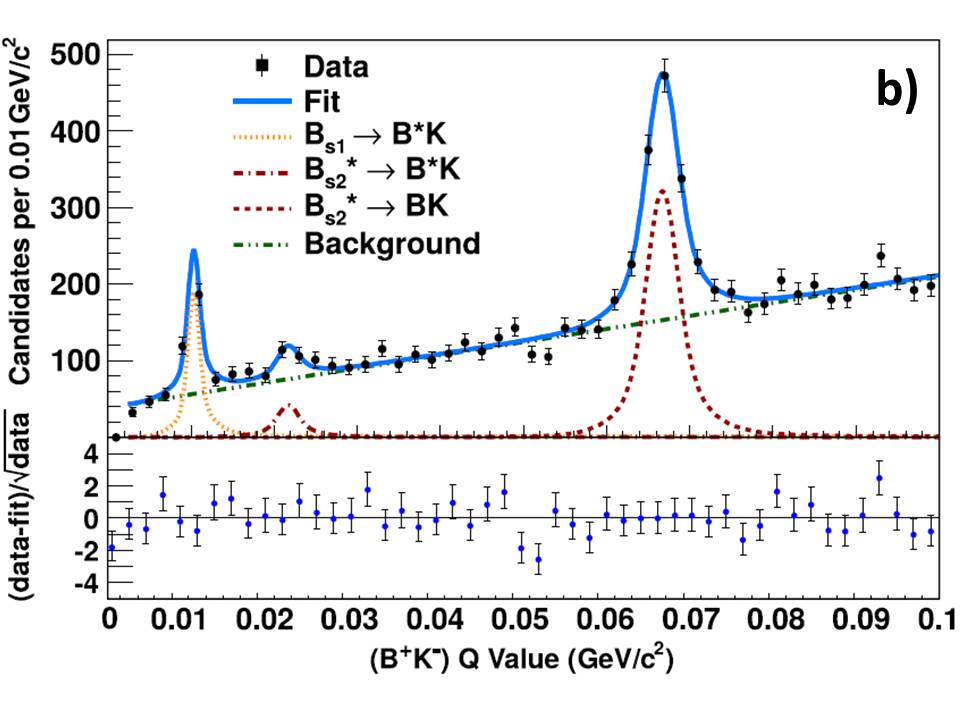} 
\end{tabular}
\caption{ {\footnotesize a): Spectrum of the allowed decays for the lowest orbitally excited B-states; 
b): Q value of the excited B-mesons: sum of all decay modes with a kaon.}}
\label{fig:fig2}
\end{center}
\end{figure}  

The latest CDF updates for the masses of the observed excited B-mesons are listed in table 
\ref{tab:tab1} and compared with most recent results from other experiments and  with the predictions 
of some HQET models.           
\begin{table}[htb]
\begin{center}
\begin{tabular}{|l|c|c|c|c|c|c|c|} 
\hline 
 & {\footnotesize $B^{0}_{1}$} & {\footnotesize $B^{+}_{1}$} & {\footnotesize $B^{*0}_{2}$} & {\footnotesize 
$B^{*+}_{2}$} & {\footnotesize $B^{0}_{s1}$} & {\footnotesize $B^{*0}_{s2}$} & {\footnotesize  Ref.} \\ 
\hline
{\footnotesize CDF} & {\footnotesize 5726.4 $\pm 1.6$} &  {\footnotesize 5726 $\pm 5$} & {\footnotesize 5736.6 
$\pm 1.7$} & 
{\footnotesize 5737.1 $\pm 1.4$} & {\footnotesize 5828.3 $\pm 0.4$} & {\footnotesize 5839.7 $\pm 0.2 $} & 
{\footnotesize \cite{aag:cdf-excited}} \\
{\footnotesize HQET} & {\footnotesize 5720} & {\footnotesize 5720} & {\footnotesize 5737} & {\footnotesize 5737} & 
{\footnotesize 5831} & {\footnotesize 5847} & {\footnotesize \cite{aah:theo1}}\\
{\footnotesize HQET} & {\footnotesize 5719} & {\footnotesize 5719} & {\footnotesize 5733} & {\footnotesize 5733} & 
{\footnotesize 5831} & {\footnotesize 5844} & {\footnotesize \cite{aai:theo2}}\\
\hline
\end{tabular}
\caption{ {\footnotesize Excited B-meson masses; the total error on the CDF results includes statistical and 
systematic errors, and the uncertainty on the B-meson masses.}}
\label{tab:tab1}
\end{center}
\end{table}

A broad structure is visible at Q value around 550 MeV/c$^{2}$ in both $B^{**0}$ and $B^{**+}$ invariant mass 
distributions (fig. \ref{fig:fig6}); the properties of the previously unobserved resonances are
measured for both neutral and charged final states; assuming a decay through the $B\pi$ channel, 
the mass values $m(B(5970)^{0}) = 5978 \pm 5 \pm 12$ and $m(B(5970)^{+}) = 5961 \pm 5 \pm 12$ are 
obtained with individual significances of $4.2\sigma$ and $3.7\sigma$ respectively.
\begin{figure}[htb]
\begin{center}
\begin{tabular}{cc}
\includegraphics[height=1.7in]{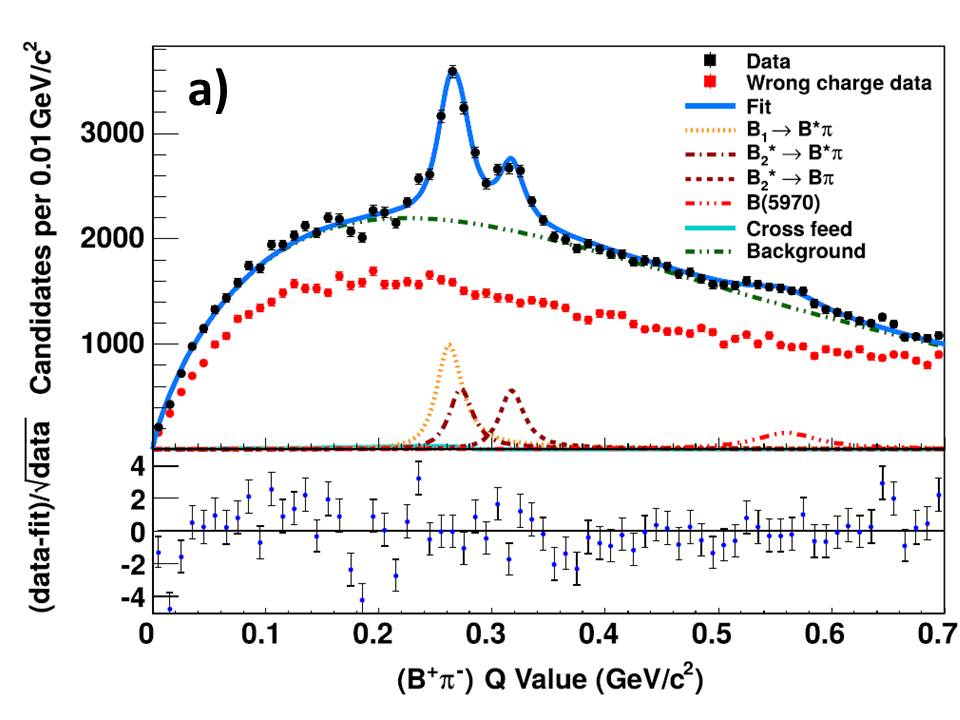} &
\includegraphics[height=1.7in]{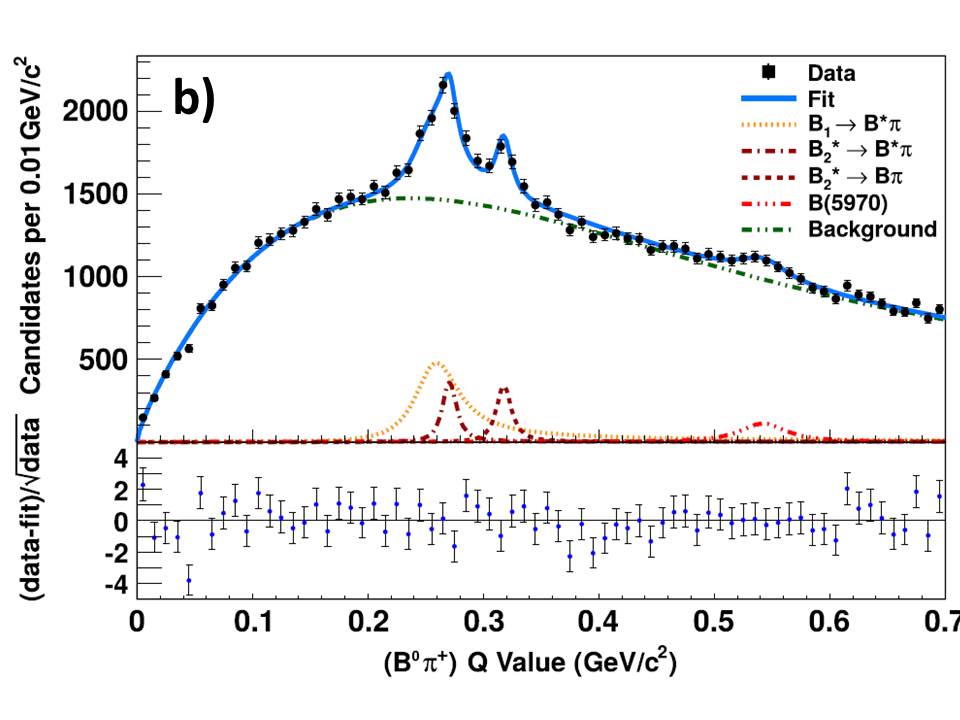} 
\end{tabular}
\caption{ {\footnotesize Distribution of the Q values of $B^{**0}$ candidates ( a) ) and $B^{**+}$ candidates 
( b) )} }
\label{fig:fig6}
\end{center}
\end{figure} 

\section{b-baryons: $\Xi^{0,-}_{b}, \Omega^{-}_{b}$}

The study of the properties of the b-baryons has been for a long time totally a Tevtaron field; many 
elements of the SU(4) (u,d,s,b) symmetry such as the $\Sigma^{(*)+}, \Sigma^{(*)-}, \Xi^{-}_{b}, 
\Omega^{-}_{b}$, and $\Xi^{0}_{b}$ baryons were first observed during the Tevatron Run II from 2006 to 
2011. The b-baryons are reconstructed at the Tevatron experiments in the $J/\Psi$ and fully hadronic modes; 
very recently, CDF made the first observation of the $\Omega^{-}_{b}$ in the fully hadronic mode;  
the invariant mass spectra of the $\Omega^{-}_{b}$ and 
$\Xi^{-}_{b}$ baryons are shown in figures \ref{fig:fig3}a) and \ref{fig:fig3}b) respectively.

After more than two years from the collision end, Tevatron results on b-baryon properties are still 
almost competitive with the first LHC results; a comparison of CDF and LHCb recent results is
shown in table \ref{tab:tab2}.
\begin{figure}[!htbp]
\begin{center}
\begin{tabular}{ccc}
\includegraphics[height=1.7in]{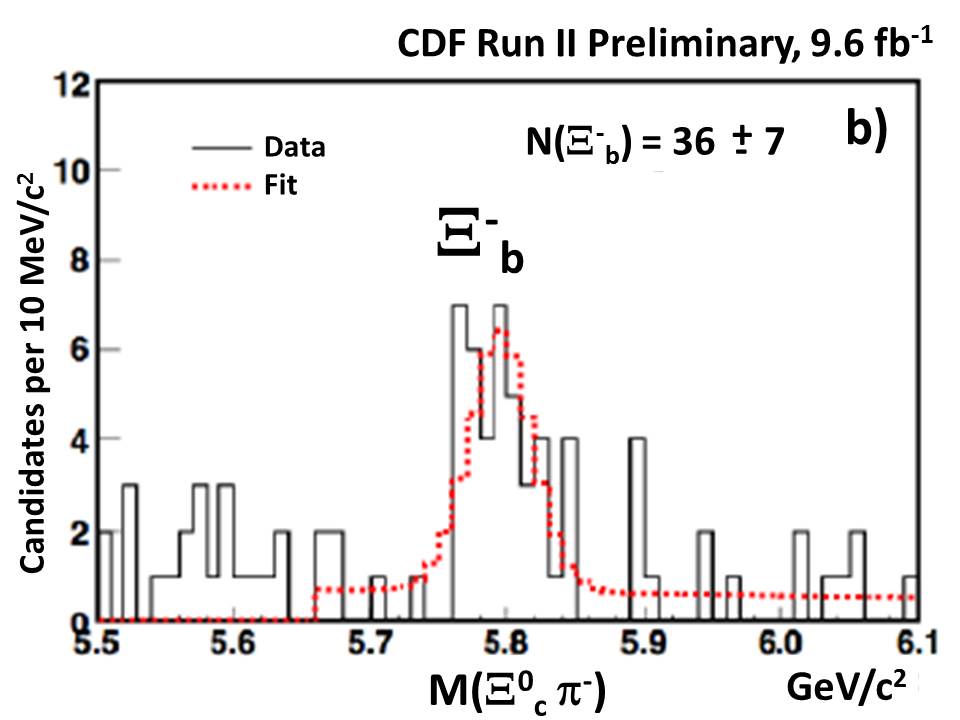} &
\includegraphics[height=1.7in]{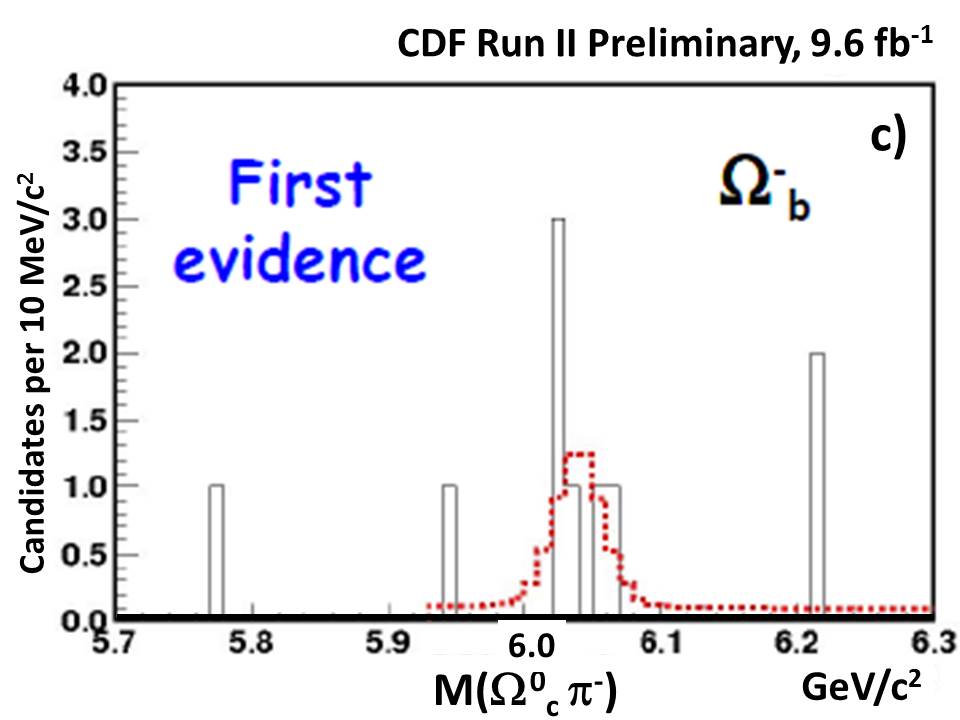}
\end{tabular}
\caption{ {\footnotesize 
a): M($\Xi^{0}_{c} \pi^{-}$) invariant mass; b): M($\Omega^{0}_{c} \pi^{-}$) invariant mass.} }
\label{fig:fig3}
\end{center}
\end{figure}  
\begin{table}[!htbp]
\begin{center}
\begin{tabular}{|c|c|c|c|c|}
\hline
& \multicolumn{2}{|c|}{ {\footnotesize CDF (ref. \cite{xyy:cdf-baryon})}} & \multicolumn{2}{|c|}{ {\footnotesize 
LHCb (ref. \cite{aaf:lhcb-baryon})}} \\
& {\footnotesize Mass (MeV/c$^{2}$)} & {\footnotesize Lifetime (ps)} & {\footnotesize Mass (MeV/c$^{2}$)} & 
{\footnotesize Lifetime (ps)} \\
\hline 
{\footnotesize $\Lambda_{b}$} & {\footnotesize 5620.15 $\pm 0.31 \pm 0.47$} & {\footnotesize  1.565 $\pm 0.035 
\pm 0.020 $} & {\footnotesize 5619.53 $\pm 0.13 \pm 0.45$} & {\footnotesize 1.482 $\pm 0.18 \pm 0.12 $} \\
{\footnotesize $\Xi^{-}_{b}$} & {\footnotesize 5793.4 $\pm 1.8 \pm 0.7$} &  {\footnotesize 1.32 $\pm 0.14 
\pm 0.02 $} & {\footnotesize 5795.8 $\pm 0.9 \pm 0.4$} & {\footnotesize 1.55 $^{+0.10}_{-0.09} \pm 0.03 $} \\
{\footnotesize $\Xi^{0}_{b}$} & {\footnotesize 5788.7 $\pm 4.3 \pm 1.4$} &  &  & \\
{\footnotesize $\Omega^{-}_{b}$} & {\footnotesize 6047.5 $\pm 3.8 \pm 0.6$} & {\footnotesize 1.66 
$^{+0.53}_{-0.40} \pm 0.02 $} & {\footnotesize 6046.0 $\pm 2.2 \pm 0.4$ }& {\footnotesize 1.54 
$^{+0.26}_{-0.21} \pm 0.05 $} \\
\hline
\end{tabular}
\caption{ {\footnotesize Summary of the latest results on b-baryon properties.}}
\label{tab:tab2}
\end{center}
\end{table}

\section{Exotic resonances: X(4140)}

The study of narrow exotic resonances in the B decay product spectrum is important to infer on possible
colorless bound quark states other than mesons and baryons; there are no theoretical reasons to exclude
meson molecules, tetra-quark aggregates, or quark-gluon hybrids, but no definitive experimental evidence
for any such states has been yet established.

The D0 experiment has recently studied the resonances in the $J/\Psi\phi$ system produced near
threshold in the decay $B^{+} \rightarrow J/ \Psi \phi K^{+}$ (and charge conjugate) \cite{aaa:D0-X4140}.
As shown in figure \ref{fig:fig4}a), a 3.1$\sigma$ evidence is obtained for the X(4140) resonance 
with mass $M_{X(4140)} = 4159 \pm 4.3(stat.) \pm 6.6 (syst.) ~MeV/c^{2}$ and width $\Gamma_{X(4140)} =
19.9 \pm 12.6(stat.) \pm 8(syst.) ~MeV/c^{2}$ in agreement with the CDF first evidence and updated 
results \cite{aab:CDF-X4140}, and with the CMS observation \cite{aac:CMS-X4140}. Debate on the 
existence of the narrow X(4140) resonance is not yet closed because of the non-observation results 
from Belle \cite{aad:Belle-X4140} and LHCb \cite{aae:LHCB-X4140}. The interpretation of the observed
resonance is also controversial; conventional charmonium should predominantly decay into $D\bar{D}$ 
pairs (not seen) with expected mass close to 3740 $MeV/c^{2}$, the open charm threshold, while the
mode $(c\bar{c}) \rightarrow J/\Psi ~hadrons$ (e.g $\phi \rightarrow KK$) is OZI suppressed. 

The summary of the experimental results on the search for narrow resonances in the $J/\Psi\phi$ 
invariant mass spectrum is shown in figure \ref{fig:fig4}b); inconsistent results are found for
the resonance around 4300 $MeV/c^{2}$.    
\begin{figure}[htb]
\begin{center}
\begin{tabular}{cc}
\includegraphics[height=1.7in]{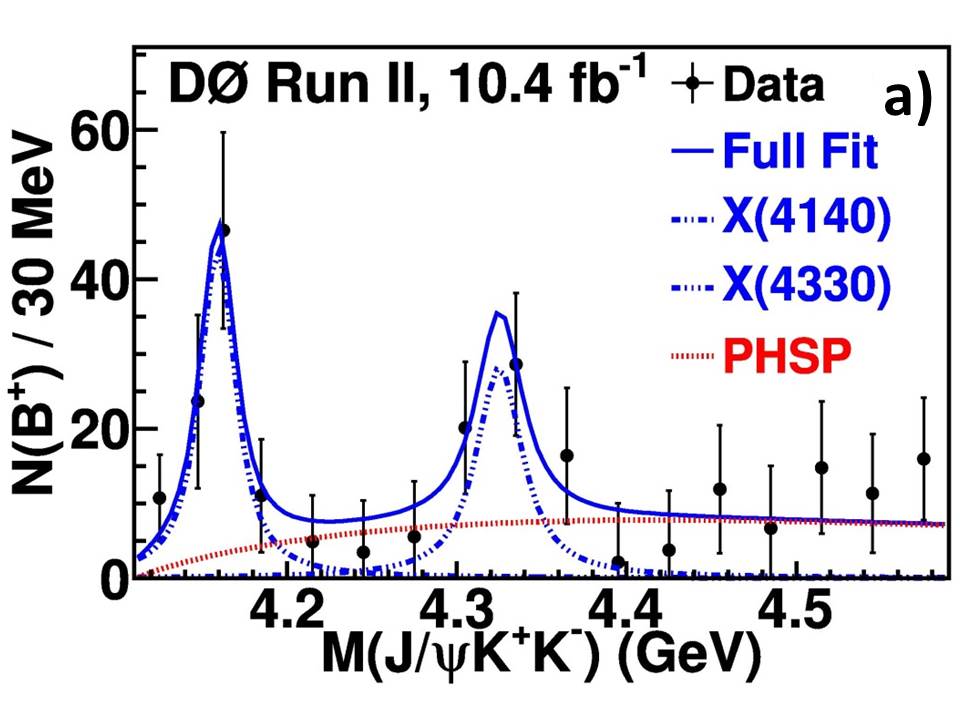} &
\includegraphics[height=1.7in]{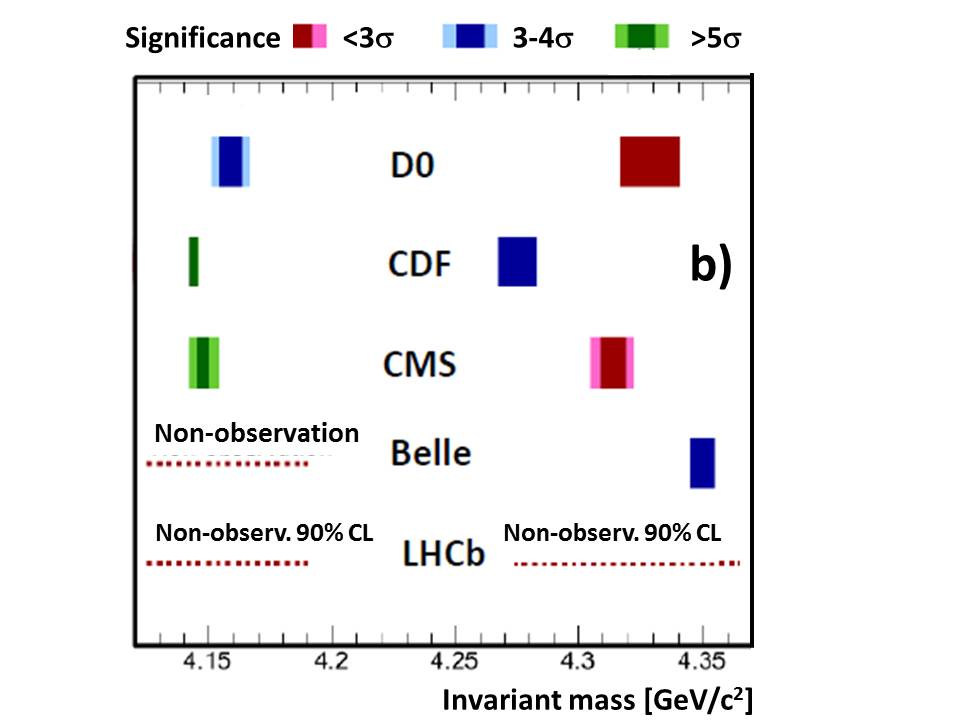} 
\end{tabular}
\caption{ {\footnotesize a): red dotted line is the fit result assuming three-body phase-space decays with no 
resonances; Breit-Wigner functions are user for the signal fit; b): summary of the resonance searches 
in the $J/\Psi \phi$ system.} }
\label{fig:fig4}
\end{center}
\end{figure}  

\section{Conclusions}

Tevatron experiments produced high quality results in heavy flavor physics during the last two decades; 
the results have been complementary and competitive with the B-Factories, showing that precision 
measurements on heavy flavor physics are possible at the hadron colliders.

Many tools and methods were developed for a clean identification of events with b-hadron production;
a rich legacy is left to LHC and to the future colliders and B-Factories.

The analysis of the full statistics samples collected by CDF and D0 is not yet completed; possible 
interesting results could be still obtained.  



\end{normalsize}

\begin{footnotesize}

\end{footnotesize}
 
\end{document}